# Research on an improved Conformer end-to-end Speech Recognition Model with R-Drop Structure

Weidong Ji[1], Shijie Zan[1], Guohui Zhou[1], and Xu Wang[1,*]

[1] College of Computer Information Engineering, Harbin Normal University, Harbin, 150025, China
[*]Corresponding Author: Xu Wang. Email: wx970125@163.com

**Abstract**：To address the issue of poor generalization ability in end-to-end speech recognition models within deep learning, this study proposes a new Conformer-based speech recognition model called "Conformer-R" that incorporates the R-drop structure. This model combines the Conformer model, which has shown promising results in speech recognition, with the R-drop structure. By doing so, the model is able to effectively model both local and global speech information while also reducing overfitting through the use of the R-drop structure. This enhances the model's ability to generalize and improves overall recognition efficiency. The model was first pre-trained on the Aishell1 and Wenetspeech datasets for general domain adaptation, and subsequently fine-tuned on computer-related audio data. Comparison tests with classic models such as LAS and Wenet were performed on the same test set, demonstrating the Conformer-R model's ability to effectively improve generalization.

**Key Words**：Speech recognition; R-drop; Conformer

## 1 Introduction

The end-to-end speech recognition process refers to the integration of acoustic, pronunciation, and language factors within a single deep neural network (DNN), simplifying the modeling process and enabling direct mapping from the input to target end [1,2]. With the wide application of deep learning across various fields, end-to-end speech recognition models have shown significantly better recognition performance than traditional statistical and machine learning models, garnering increasing attention [3-5].

Currently, there are several research directions in the field of end-to-end speech recognition: (1) the Connectionist Temporal Classification (CTC) method [6,7], which uses forward algorithm and backward algorithm to compute the current model loss and optimize it [8-10]. Lee et al. further proposed the intermediate CTC loss to regularize CTC training and improve performance [11]. (2) The Recurrent Neural Network Transducer (RNN-T) [12,13], which is used for streaming speech recognition and can memorize forward information. (3) The Encoder-Decoder architecture based on the attention mechanism [14-16], which has made significant progress in recent years. Chan et al. proposed the Listen, Attend, Spell (LAS) model [17], which integrates acoustic, pronunciation, and language factors for end-to-end speech recognition. Miao et al. [18] studied online streaming speech recognition by adding a window to the Transformer.Han et al. proposed the ContextNet model [19], which achieves good recognition performance while reducing model parameters through progressive downsampling and model scaling. The U2++ architecture in the Wenet framework developed by the Northwestern Polytechnical University's Speech Laboratory [20,21] combines CTC with a Transformer decoder for superior recognition accuracy. Liu et al. replaced all audio preprocessing steps with a neural network, reducing the impact of preprocessing accuracy on subsequent steps [22]. Fu et al. applied supervised contrastive learning in the proposed SCaLa model to enhance phoneme-level representation learning [23]. Peng et al. achieved excellent recognition performance by combining self-attention mechanism with multi-layer perceptron module into two branches in their model [24]. Maekaku et al. improved the Transformer model using attention weight smoothing, resulting in increased accuracy [25]. In summary, end-to-end speech recognition models generally have higher accuracy than traditional models. However, further analysis reveals that these models tend to ignore the evaluation index of generalization. Different models are suitable for different datasets, i.e., achieving high accuracy on a specific dataset but with a large decrease in accuracy when applied to other datasets. Therefore, these models generally have generalization issues, and the accuracy of applying a trained model to new speech data needs to be improved.

We propose the Conformer-R model to address these issues. In addition to applying convolutional operations to the Encoder layer of the Transformer and using it as the encoding layer of our proposed Conformer-R model, we also utilize the R-Drop structure as a crucial component to enhance the model's generalization ability. R-Drop forces the output distributions of different sub-models generated by dropout to be consistent with each other, thus reducing overfitting and improving the model's ability to generalize.

## 2 Related Work

The Transformer is a model proposed by Google in 2017 for sequence-to-sequence tasks, originally used for machine translation and now widely applied in various deep learning fields. It consists of two parts: the Encoder responsible for feature extraction and the Decoder responsible for decoding and completing specific tasks, as shown in Figure 1: The advantage of the Transformer is that it does not require prior design of input-target correspondence rules, but automatically learns the correspondence rules in a "black-box" form. It is also effective in dealing with significant differences in the lengths of input and target sequences.

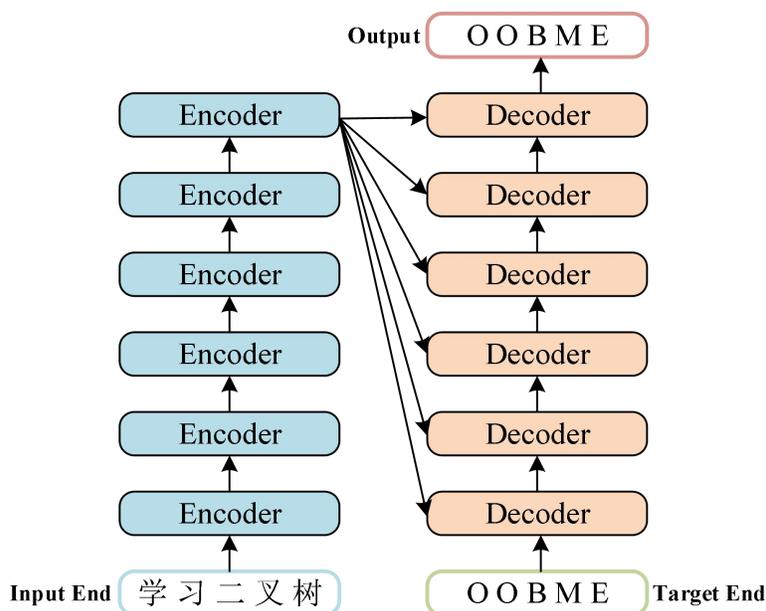

**Figure 1:** Encoders and decoders in Transformer

The detailed structure of the Transformer encoder-decoder is illustrated in Figure 2. The input data is first encoded and fused through the embedding layer, which includes the encoding of input data and absolute position. Then, it is encoded by the encoder with the self-attention mechanism. The encoded result is used as the key-value vector of the decoder's self-attention mechanism. Finally, the decoder decodes the target sequence through an auto-regressive manner.

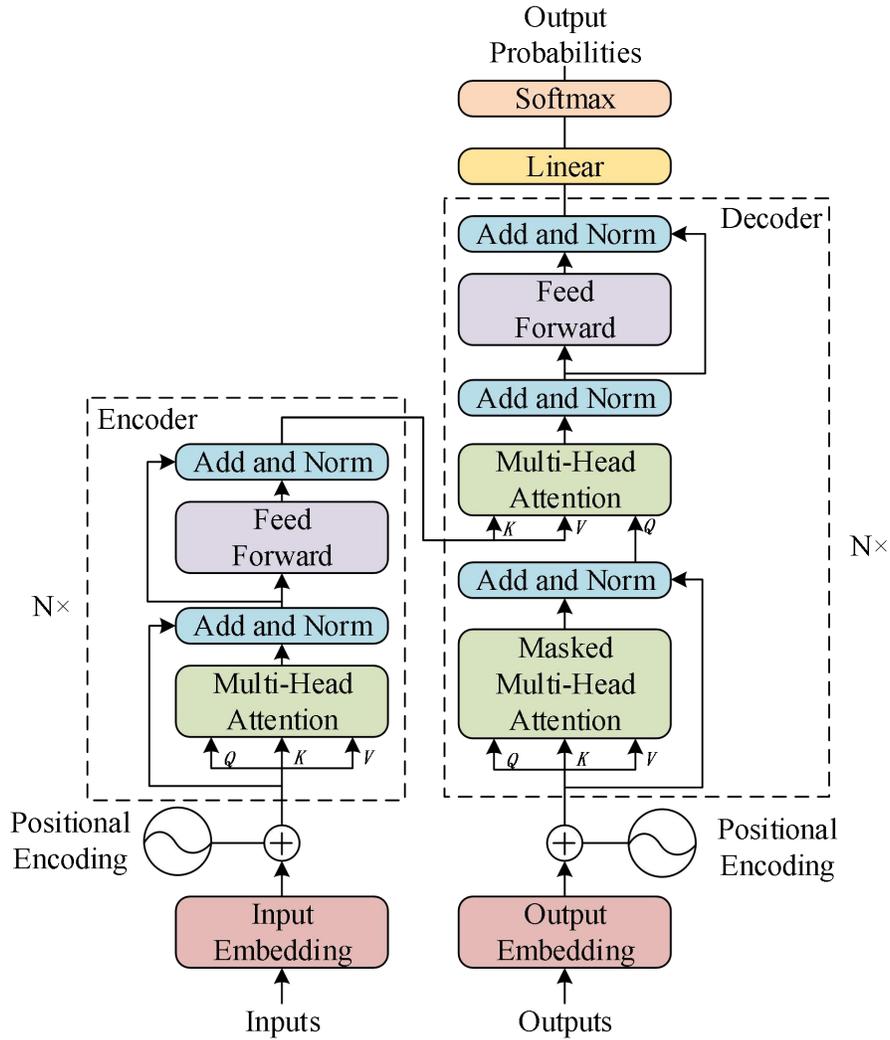

**Figure 2:** Detail structure of Transformer

The Transformer model uses multi-head self-attention mechanism to extract features of input vectors, which may cause the model to lose positional information of long sequence inputs. Therefore, the Transformer introduces absolute positional encoding for each word vector, generating unique texture coding information. By adding each word vector with its absolute positional encoding, the Transformer can learn the temporal dependency of input data.

The embedded vectors after positional encoding need to undergo the operation of multi-head self-attention mechanism. The input matrix is duplicated as query matrix (Q matrix), key matrix (K matrix) and value matrix (V matrix), and then each matrix is fed into its own linear layer for computation and output, as shown in Figure 3.

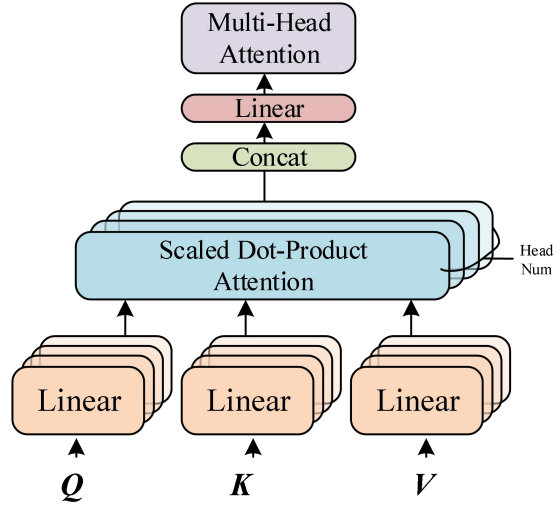

**Figure 3:** Multi-head self-attention mechanism

The scaled dot-product attention mechanism performs matrix multiplication of the linearly transformed Q, K, and V matrices, resulting in the attention scores $A_{ij}$ for input variables *i* and *j*. The weighted sum of the V matrix is then computed based on the attention scores to obtain the output $Attention(A_{ij}, V)$ of the self-attention mechanism, as shown in equations (1) and (2).

$$A_{ij} = QK^T \tag{1}$$

$$Attention(A_{ij}, V) = softmax(\frac{A_{ij}}{\sqrt{d_k}})V \tag{2}$$

The output of the multi-head self-attention mechanism $MultiHead(Q,K,V)$ can be obtained by concatenating the results of multiple heads, as shown in equation (3).

$$MultiHead(Q,K,V) = Concat(head_1,...,head_h)W^o \tag{3}$$

After encoding the input information, the encoded matrix is used as the key-value matrix input to the decoder, waiting for the target decoding.

The decoder of Transformer is an auto-regressive decoder, and therefore adopts a masked multi-head attention mechanism. During the auto-regressive decoding process, the current word embedding at a position can only access the word embeddings before it in the temporal sequence, and cannot access the information of word embeddings after it. A lower triangular mask matrix is usually applied to mask the target side of the training data, as shown in Figure 4. Then, the masked target embedding matrix is subjected to multi-head self-attention operation to obtain the decoding result.

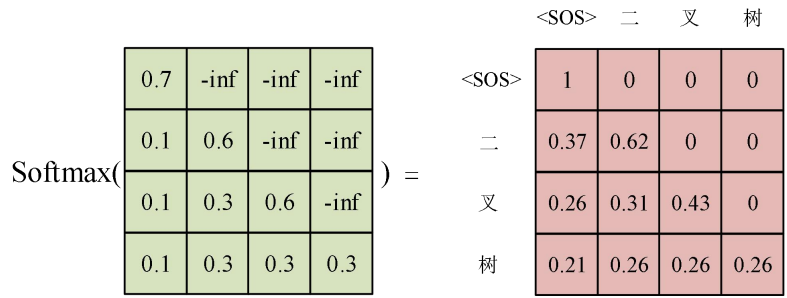

**Figure 4:** Masked multi-head attention mechanism

A typical Sequence-to-Sequence model compresses the preceding context information into a fixed-length vector, which performs poorly in decoding long input sequences. Transformer addresses this issue by encoding all information into a matrix via multi-head self-attention mechanism, effectively resolving the problem of long sequences and associating the input and target sequences, thus improving the decoding accuracy of the target side.

## 3 Conformer-R Model

By integrating the R-Drop structure with the Conformer model, the Conformer-R model is constructed to enhance the generalization ability of the Conformer model. The structure of the Conformer-R model is shown in Figure 5.

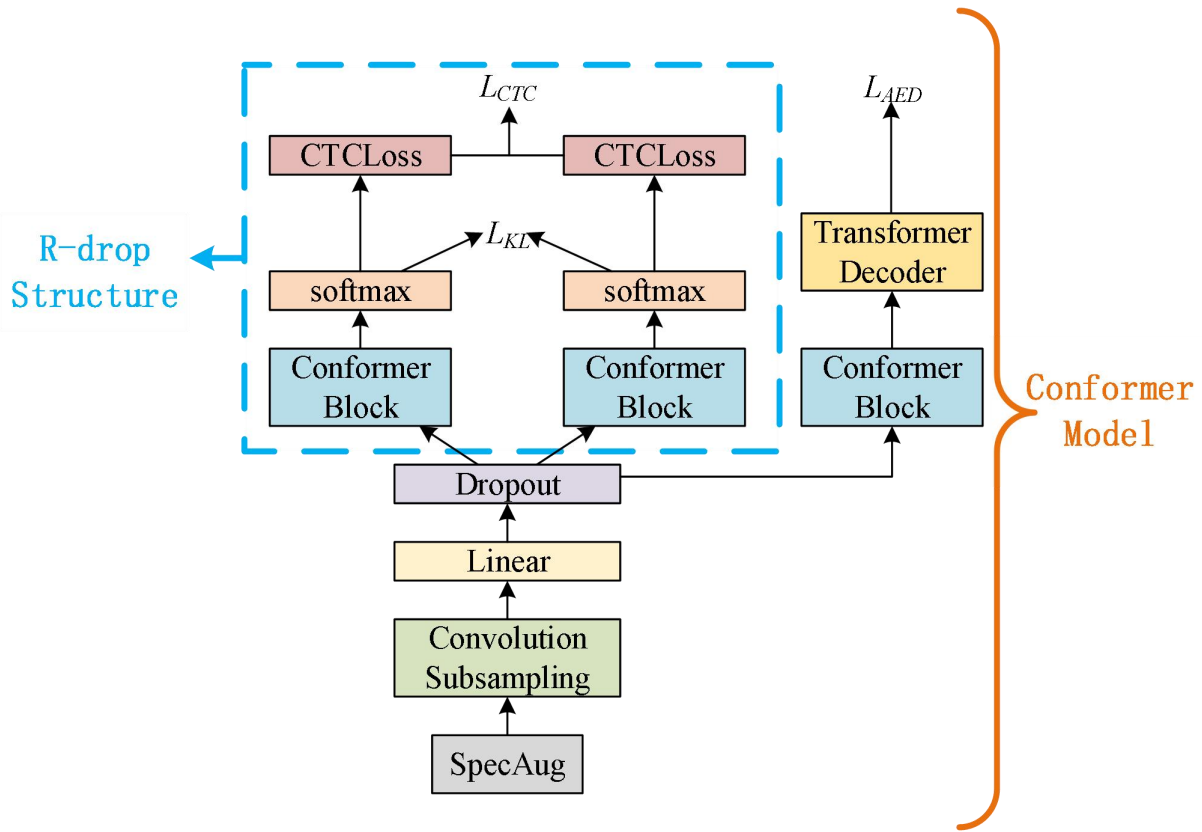

**Figure 5:** Conformer-R model

In the Conformer-R model, data is passed through two Conformer-blocks with different dropout rates, resulting in two different distributions. The KL divergence loss $L_{KL}$ is calculated between the output distributions, followed by the CTC loss $L_{merge}$. The total loss $L_{CTC}$ is then obtained by combining the two losses with a weighted sum, as shown in Equation (4).

$$L_{CTC} = (1-\alpha)L_{merge} + \alpha L_{KL} \tag{4}$$

During the computation of the CTC loss, the output vectors are independent of each other, and the model does not consider their dependency. The Conformer model utilizes a Transformer decoder to decode the input data, and during training, the model performs self-attention mechanism operations on the target context. This is equivalent to training a language model that makes the output distribution not independent, thus learning the dependency between target contexts.

The Transformer decoder calculates the cross-entropy loss function of the target side, obtaining the attention-based encoder-decoder (AED) loss. The CTC loss and AED loss are weighted and combined to obtain the final loss function, as shown in equation (5).

$$L = (1-\beta)L_{CTC} + \beta L_{AED} \tag{5}$$

The Conformer-R model integrates the R-Drop structure on top of the Conformer model, enhancing the model's generalization ability. In this model, the propagation path of CTC loss is much shorter than that of AED loss, which overcomes the issue of the error backpropagation path being too long for Conformer blocks in AED.

### 3.1 Conformer Speech Recognition Model

Conformer is an end-to-end speech recognition model proposed by Google in 2020 [26], which is an improvement on the Transformer model. The model combines the global modeling capability of attention mechanism with the local modeling capability of convolutional neural networks. It compresses the length of speech sequences through convolutional operations while improving the feature extraction effectiveness of the model in both global and local contexts. The basic structure of Conformer is shown in Figure 6.

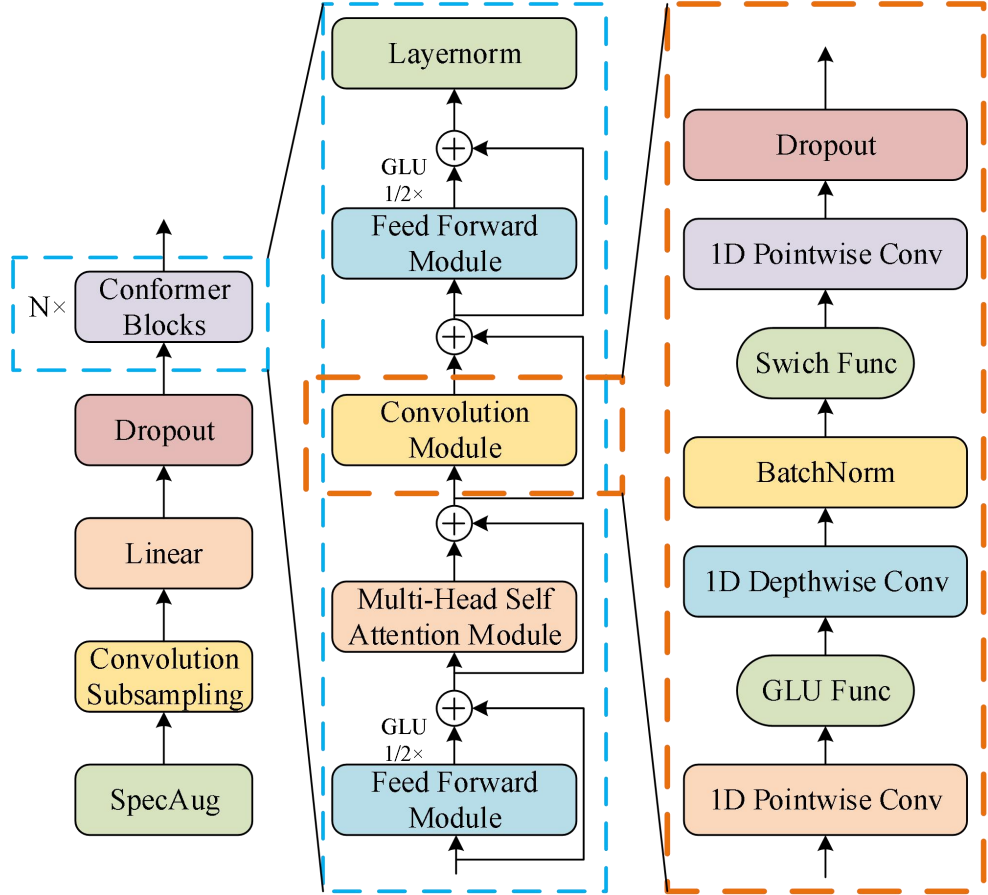

**Figure 6:** Structure of Conformer

The encoder of the Conformer model first reduces the dimensionality of the input data using a Convolution Subsampling layer and then processes it using multiple Conformer blocks. The model's notable feature is the use of Conformer blocks instead of Transformer blocks. A Conformer block consists of four modules stacked together, namely a feed-forward module, a multi-head self-attention module, a convolution module, and a second feed-forward module. This approach has been proven to address the issue of lengthy speech signals and is effective on the LibriSpeech test set [26].

*3.1.1 Sinusoidal Relative Positional Encoding*

The position encoding in the Multi-Headed Self-Attention (MHSA) module of Conformer is different from the absolute position encoding used in Transformer. Conformer uses the relative sinusoidal position encoding introduced in Transformer-XL [27], which can improve the performance of the self-attention module on variable-length input and enhance the robustness of the model to variable-length speech signals. In Transformer, the word embedding vectors and absolute position encoding of input vectors $i$ and $l$ are combined to calculate the attention score $A_{il}$ using the formula (1). By separating and expanding the word embedding vectors and absolute position encoding, we can obtain the formula (6).

$$A_{il} = e_i \mathbf{W}^Q (\mathbf{W}^K)^T e_l^T + e_i \mathbf{W}^Q (\mathbf{W}^K)^T p_l^T + p_i \mathbf{W}^Q (\mathbf{W}^K)^T e_l^T + p_i \mathbf{W}^Q (\mathbf{W}^K)^T p_l^T \tag{6}$$

Where $e_i$ and $e_l$ are the word embedding vectors for $i$ and $l$ respectively, and $p_i$ and $p_l$ are the absolute position encodings for $i$ and $l$. Furthermore, by changing the absolute position encodings in the equation to relative position encodings, the attention score calculation formula used in the Conformer model can be obtained, as shown in equation (7).

$$A_{il} = e_i \mathbf{W}^Q (\mathbf{W}^{K,E})^T e_l^T + e_i \mathbf{W}^Q (\mathbf{W}^{K,r})^T r_{i-l}^T + u(\mathbf{W}^{K,E})^T e_l^T + v(\mathbf{W}^{K,r})^T r_{i-l}^T \tag{7}$$

For the third and fourth terms in Equation (7), since the absolute position encoding of i is not needed when considering relative position encoding, $p_i W^Q$ needs to be encoded as learnable vectors $u$ and $v$, while $p_l$ is directly converted to the relative position encoding $r_{i-l}$. $W^K$ is transformed to $W^{K,E}$ and $W^{K,r}$ as the key weight matrices with contextual information and relative position information, respectively.

*3.1.2 The Convolution Module*

The Conformer model uses a linear gated unit (GLU) [28] before the convolution module to enhance the model's generalization ability, followed by a 1-D deep convolution layer. Batch normalization is deployed after convolution to help train deep models. Finally, the Swish activation function and dropout are applied to regularize the network. The internal operation of the convolution module is shown in Equation (8), where $h_l(x)$ is the output of the convolution module, $w$, $b$, $v$, and $c$ are learnable parameters, and $\sigma$ is the sigmoid function.

$$h_l(x) = (x * w + b) \otimes \sigma(x * v + c) \tag{8}$$

The internal configuration of the convolutional module is shown in Figure 7.

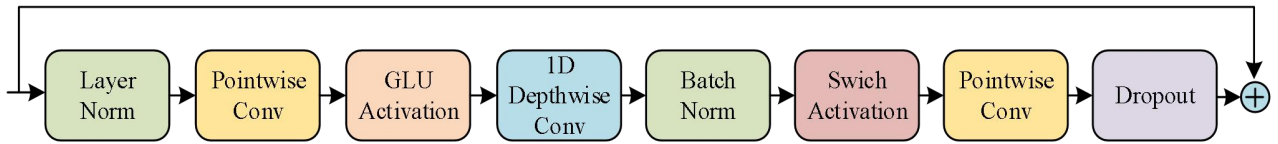

**Figure 7:** Convolution module of Conformer

*3.2 R-Drop Structure*

In practical engineering, the biggest challenge faced during training of speech recognition models is their low generalization ability due to the limited information contained in speech signals compared to richer modalities such as images and videos, leading to overfitting on small datasets such as the TIMIT dataset. To address this issue, this paper introduces the R-Drop structure [29] into the Conformer architecture to enhance the model's generalization ability. The R-Drop structure is shown on the right side of Figure 8.

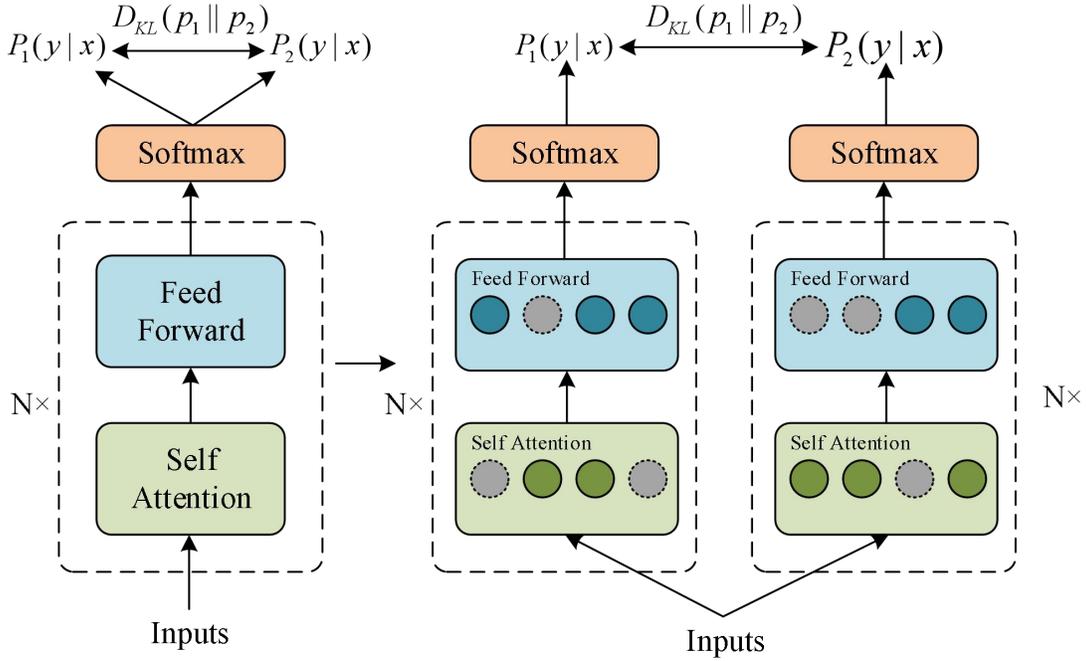

**Figure 8:** R-Drop structure

A combined model with R-Drop structure inputs each training sample into the same model twice. Due to the random unit deletion in the Dropout operation, the two paths can be approximated as different models, resulting in two different distributions $P_1^\theta(y_i|x_i)$ and $P_2^\theta(y_i|x_i)$ predicted by the model. As shown on the right side of Figure 8, the units removed by Dropout in each layer of the left path are different from those in the right path. The two outputs obtained by R-Drop are first used to calculate the cross-entropy loss $L_i^{CE}$, as shown in equation (9), and then the bidirectional Kullback-Leibler (KL) divergence is calculated to obtain the model KL divergence $L_i^{KL}$, as shown in equation (10).

$$L_i^{CE} = -\log P_1^\theta(y_i|x_i) - \log P_2^\theta(y_i|x_i) \tag{9}$$

$$L_i^{KL} = \frac{1}{2}[KL(P_1^\theta(y_i|x_i) \| P_2^\theta(y_i|x_i)) + KL(P_2^\theta(y_i|x_i) \| P_1^\theta(y_i|x_i))] \tag{10}$$

To ensure that the outputs produced by the model using different Dropouts are as consistent as possible, the bidirectional KL divergence value is used as a term in the loss function and minimized. Finally, the two losses are weighted and combined to obtain the overall loss, as shown in Equation (11).

$$L_i = L_i^{CE} + \alpha L_i^{KL} \tag{11}$$

The variable $\alpha$ controls the weight coefficient of $L_i^{KL}$, and experimental results show that the optimal value of $\alpha$ is usually between 1 and 10, depending on the specific task.

During the actual training process, the approach used is not to input the same data into the same model twice. Instead, in order to save training time, the training data is usually directly copied to double the batch size. Although this increases the model complexity and training time per step, it can save the total training time.

## 4 Experimental Design and Result Analysis

### 4.1 WER and CER

In the evaluation of English speech recognition models, Word Error Rate (WER) is commonly used to assess the model performance. However, for Chinese speech recognition models, Character Error Rate (CER) is usually used to measure the model performance. The calculation of CER involves the Levenshtein distance between two strings, and the formula is shown in Equation (12).

$$CER = (S+D+I)/N = (S+D+I)/(S+D+H) \tag{12}$$

Where $S$ represents the number of substituted characters, $D$ represents the number of deleted characters, $I$ represents the number of inserted characters, $H$ represents the number of correctly recognized characters, and N represents the total number of characters in the target sequence.

The formula for calculating character accuracy is shown in Equation (13), which represents the accuracy of the recognition results.

$$CER.acc = 1 - CER = (N-D-S-I)/N = (H-I)/N \tag{13}$$

### 4.2 Datasets and Parameters

This experiment used the wenetspeech[30] 10,000-hour Chinese speech recognition training set from Northwestern Polytechnical University and the Aishell1[31] dataset from Shell Research to pre-train a general-purpose speech recognition model. The wenetspeech dataset contains over 10,000 hours of high-quality labeled data from a variety of speech scenarios, topics, fields, and noise environments, sourced from YouTube and podcast videos. The YouTube data labels were obtained by OCR recognition of video subtitles, while the podcast data labels were obtained using a high-quality speech recognition transcription system. Additionally, wenetspeech provides a Dev set for cross-validation and two Test sets (Test-net and Test-meeting). The Test-net dataset is sourced from the internet and covers popular fields such as game commentary and live streaming, while the Test-meeting dataset is sourced from 197 real meetings and covers fields such as education, real estate, and finance, in a far-field noise environment. The Aishell1 dataset includes a 150-hour training set from 340 speakers, a 10-hour validation set from 40 speakers, and a 5-hour test set from 20 speakers, covering music, TV dramas, news, and other sources.

In all experiments, the model extracted 80-dimensional Fbank speech features as input. The total dictionary of all training sets contained 4,419 characters. The encoder consisted of a R-Drop two-branch stacked 8-layer Conformer block, while the decoder consisted of a stacked 4-layer Transformer decoder. This experiment used an Adam optimizer with a warm-up learning rate mechanism for model training, with Adam parameter $\beta_1 = 0.9$, $\beta_2 = 0.98$, $\varepsilon=10^{-9}$, and a dynamically adjusted learning rate calculated using Equation (14).

$$lr = k * d_m^{-0.5} \min(step^{-0.5}, step * warmup\_steps^{-1.5}) \tag{14}$$

Where $k=1$, $d_m$ is set to 512 for the encoder output dimension, and warmup_steps is set to 12000 for the preheating steps. The dropout value for the R-Drop branch is set to 0.1, and label smoothing with a correct probability of 0.9 is used for KL divergence calculation. Batch_bins are used for batch training, with an initial value of 1500000 for the total number of feature frame elements in each batch, and a gradient accumulation of 4. All models in the experiment were trained for 25 epochs using 8 GTX 3090Ti GPUs continuously, taking about 20 days.

*4.3 Pre training Experiment and Result Analysis*

In our study, the Conformer-R model was trained and compared with other models on the wenetspeech+Aishell1 training set. The model was validated on the dev set provided by wenetspeech, and tested on the Test-meeting and Test-net test sets provided by wenetspeech, as well as the Test-ai test set provided by Aishell. The character error rate (CER) metric was calculated using the CER calculation function available on Huggingface. The experimental results are shown in Table 1.

**Table 1:** CER of different models on the wenetspeech+aishell1 dataset

| Model | Dev CER | Test-meeting CER | Test-net CER | Test-ai CER |
|---|---|---|---|---|
| LAS | 12.4% | 20.3% | 12.3% | 8.4% |
| Espnet(Transformer)[32] | 10.8% | 16.1% | 10.8% | **7.0%** |
| Espnet(BiLSTM)[32] | 12.4% | 15.4% | 11.2% | 7.3% |
| Wenet(U2++)[21] | 11.1% | 19.8% | 10.5% | 7.1% |
| LFMMI[31] | 13.2% | 19.4% | 12.4% | 9.1% |
| Conformer-R(ours) | **9.8%** | **14.9%** | **10.2%** | 7.2% |

From Table 1, it can be observed that our baseline model outperforms the Transformer encoder and BiLSTM models provided by the ESPnet toolkit, the U2++ Transformer model provided by the Wenet toolkit, and the LFMMI model provided by the Kaldi toolkit under the same training conditions. The Conformer-R model achieves a CER improvement of 1% to 5% on the test set, indicating a significant improvement in the effectiveness of the Conformer-R model in speech recognition.

We conducted ablation experiments on the Conformer-R model to verify the effectiveness of the Conformer-R strategy. The experimental results are shown in Table 2.

**Table 2:** Conformer-R model ablation experiments

| Model | Dev CER | Test-meeting CER | Test-net CER | Test-ai CER |
|---|---|---|---|---|
| Conformer(BiLSTM-Decoder) | 12.1% | 19.2% | 16.2% | 10.2% |
| Conformer-R(BiLSTM-Decoder) | 10.1% | 17.9% | 15.3% | 9.4% |
| Conformer(Transformer-Decoder) | 11.3% | 18.3% | 12.6% | 9.3% |
| **Conformer-R(Transformer-Decoder)(Ours)** | **9.8%** | **14.9%** | **10.2%** | **7.2%** |

According to the results of the ablation experiments shown in Table 2, the use of the original Conformer structure resulted in an average increase in character error rate of 2.5% compared to Conformer-R. Using a Transformer decoder as the model decoder reduced the character error rate by approximately 3%. The ablation experiment results demonstrate the effectiveness of the R-Drop structure and the Transformer decoder. The addition of these strategies can improve model performance and increase the recognition accuracy of the speech recognition model. The training loss of the Conformer-R model in the ablation experiment is shown in Figure 9, and the results shown in the figure are smoothed with a Tensorboard coefficient of 0.5.

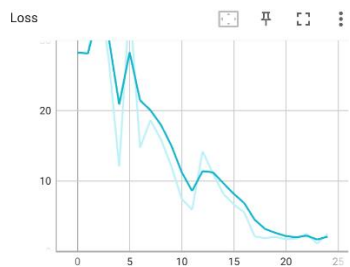

(a) Conformer-R(Transformer) Loss

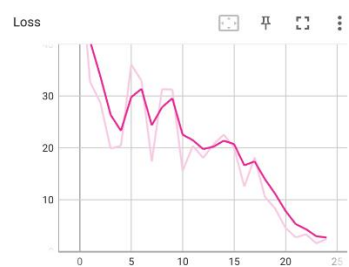

(b) Conformer-R(BiLSTM) Loss

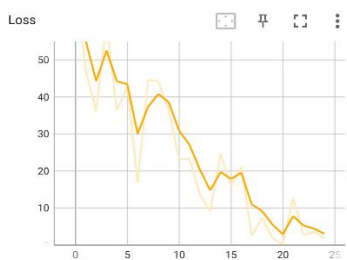

(c) Conformer(Transformer) Loss

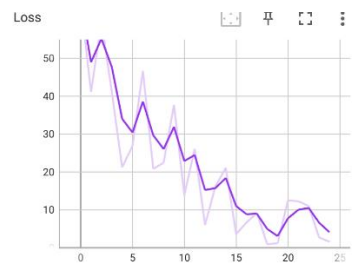

(d) Conformer(BiLSTM) Loss

**Figure 9:** Loss trend of Conformer-R model ablation experiments

From Figure 9, it can be observed that all models in the ablation experiment converge normally, indicating the validity of the ablation experiment results. Among them, the experiment combining Conformer-R model with Transformer decoder achieves the best performance, with the fastest loss reduction and little oscillation.

### *4.4 Analysis of Fine-tuning Experimental Results*

The data used for fine-tuning in this study consists of 113 hours of computer course teaching videos and corresponding subtitles with timestamps collected from the Internet. Approximately 10 hours of data were set aside as the test set. The remaining videos were segmented and aligned with subtitles to generate the fine-tuning training set. The final training set had a duration of approximately 400 hours, which was obtained by mixing the pre-training training set data and the fine-tuning training set data in a 3:1 ratio. The pre-trained Conformer-R model was fine-tuned on the vertical domain training set, with the validation set used during pre-training. The CER values obtained from the experiments are compared in Table 3.

**Table 3:** Fine tuning training CER comparison

| Model | Dev CER | Test CER |
|---|---|---|
| Conformer-R universal pre training model | 12.1% | 13.5% |
| Conformer-R vertical fine-tuning model | **11.7%** | **6.3%** |

From Table 3, it can be seen that the vertical domain fine-tuning experiment is effective. By fine-tuning the model with vertical domain data, the model learns vertical domain-specific terminology, effectively enhancing the adaptability of the speech recognition model in the vertical domain.

## 5 Conclusion and Future work

To address the issue of poor generalization ability in current speech recognition systems, a Conformer-R model with an R-Drop structure is proposed. The model fuses the R-Drop structure and re-designs the loss function to enhance the model's generalization ability through multi-path dropout. The model is pre-trained on the Aishell1 and Wenetspeech datasets for general domain pre-training and performs well on the test_meeting and test_net test sets provided by Wenetspeech, as well as the test_ai test set provided by Aishell1. The CER value in the comparative experiment is lower than that of other comparative models, and the CER value in the ablation experiment is lower than that of other ablation variants. The model is fine-tuned using a related audio training set in the computer domain, and the CER value on the domain test set decreases significantly, proving the effectiveness of fine-tuning. In the future, attempts will be made to improve the model's performance, such as using the same dropout value for different sub-models based on the R-drop structure and calculating the bidirectional KL divergence, or training two sub-models with different dropout values, respectively, as the main and auxiliary models and calculating the unidirectional KL divergence.

*Funding Statement*:

1. the National Natural Science Foundation of China(31971015)

2. Natural Science Foundation of Heilongjiang Province in 2021(LH2021F037)

*Author Contributions:*

The authors confirm contribution to the paper as follows: study conception and design: Weidong Ji, Shijie Zan; data collection: Xu Wang; analysis and interpretation of results: Weidong Ji, Shijie Zan, Guohui Zhou; draft manuscript preparation: Shijie Zan, Xu Wang. All authors reviewed the results and approved the final version of the manuscript.

**Conflicts of Interest:**

The authors declare that they have no conflicts of interest to report regarding the present study.